\documentclass[%
reprint,
superscriptaddress,
 amsmath,amssymb,
 aps,
prl,
floatfix,
]{revtex4-2}

\usepackage{graphicx}
\graphicspath{{figures/}} 
\usepackage{dcolumn}
\usepackage{bm}
\usepackage{physics}
\usepackage{tikz}
\usepackage{xcolor}
\usepackage{siunitx}
\DeclareSIUnit{\molar}{M}

\usepackage{tabularx}
\usepackage{xspace}
\usepackage{comment}
\usepackage[ruled,vlined]{algorithm2e}
\usepackage{algpseudocode}

\newcommand{\mytilde}{\raise.17ex\hbox{$\scriptstyle\mathtt{\sim}$}}
\usepackage{etoolbox}

\usepackage[colorlinks=true,allcolors=blue]{hyperref}

\newcommand \ucsb {Department of Physics, University of California at Santa Barbara, Santa Barbara, California 93106, USA}

\newcommand \ustc{Polymer Science and Engineering Department, University of Science and Technology of China, Hefei, 230022, China}

\begin{document}

\title{Rectified Rotational Dynamics of Colloidal Inclusions in Two-Dimensional Active Nematics}

\author{Sattvic Ray}
\affiliation{\ucsb}

\author{Jie Zhang}
\affiliation{\ustc}

\author{Zvonimir Dogic}
\affiliation{\ucsb}

\date{\today}

\begin{abstract}
We investigate the dynamics of mobile inclusions embedded in 2D active nematics. The interplay between the inclusion shape, the boundary-induced nematic order, and the autonomous flows powers the inclusion motion. Disks and achiral gears exhibit unbiased rotational motion, but with distinct dynamics. In comparison, chiral gear-shaped inclusions exhibit long-term rectified rotation, which is correlated with dynamics and polarization of nearby +1/2 topological defects. The chirality of defect polarities and the active nematic texture around the inclusion correlate with the inclusion's instantaneous rotation rate. Inclusions provide a promising tool for probing the rheological properties of active nematics and extracting ordered motion from the inherently chaotic motion of active nematics.  
\end{abstract}

\maketitle

Active matter is composed of energy-consuming constituents whose interactions generate large-scale dynamical patterns \cite{Marchetti2013Review}. A particular realization of these materials are active nematic liquid crystals, which are composed of motile anisotropic particles with head-tail symmetry. Active nematics exhibit autonomous chaotic flows which are driven by internally generated active stresses that are proportional to the local particle alignment \cite{Simha_Ramaswamy_2002, Ramaswamy_Menon_2007, Sanchez_Dogic_2012,Zhou2014,guillamat2016control,duclos2018spontaneous,li2019data,kumar2018tunable,tan2019topological}. Their dynamics are closely associated with the creation and annihilation of topological defects, which acquire motility from the active stresses \cite{Shi2013, Giomi2013, Thampi2014, giomi2015geometry, gao2015multiscale, Putzig2016, Shankar2018, vliegenthart2020filamentous,PhysRevLett.129.258001}. Transforming the inherently chaotic dynamics of active nematics into predetermined spatiotemporal patterns that can perform useful functions remains a challenge. One method of controlling active fluids is through boundaries and geometrical confinement. For example, static boundaries can stabilize isolated vortices and arrays of vortices of active fluids \cite{Wioland_Goldstein_2013, Nishiguchi_Sokolov_2018}. Furthermore, rigid boundaries can also generate unidirectional coherent flows on macroscopic scales and control the periodic nucleation and motion of topological defects in active nematics~\cite{Wu_Dogic_2017, Opathalage2019, Sagues2019Confinement,hardouin2020active,figueroa2022non,hardouin2022active}. In a complementary direction, active fluids can interact with rigid but motile boundaries that are advected by autonomous flows. A foundational example is the observation of the rectified rotation of a chiral gear-shaped particle powered by motile bacteria~\cite{DiLeonardo2010, Sokolov2010}. The individual bacteria interacted with the boundaries, generating asymmetric forces which powered persistent rotational motion. However, the full range of interaction mechanisms between active fluids with rigid motile inclusions remains unexplored, especially in the regime where the active fluid is a \textit{continuum} with orientational liquid crystalline order and the associated topological defects.  

To address this knowledge gap, we studied the dynamics of movable inclusions embedded in 2D microtubule-based active nematic. We observed complex interplay and feedback between the motion of a motile inclusion and the nematic texture of the enveloping liquid crystal. The inclusion boundaries control the formation and spatiotemporal patterning of topological defects. In turn, the defects and associated nematic texture exerted stress on the motile inclusion producing persistent rotations. When combined with work which explored how active fluids couple to deformable liquid interfaces and membranes ~\cite{keber2014topology,vutukuri2020active,takatori2020active,kokot2022spontaneous, Adkins2022,young2021many}, our experiments demonstrate the need to develop a unified theoretical framework for quantitative understanding of interactions between active fluids and movable, deformable and reconfigurable boundaries.

\begin{figure}
    \centering
    \includegraphics[width=\columnwidth]{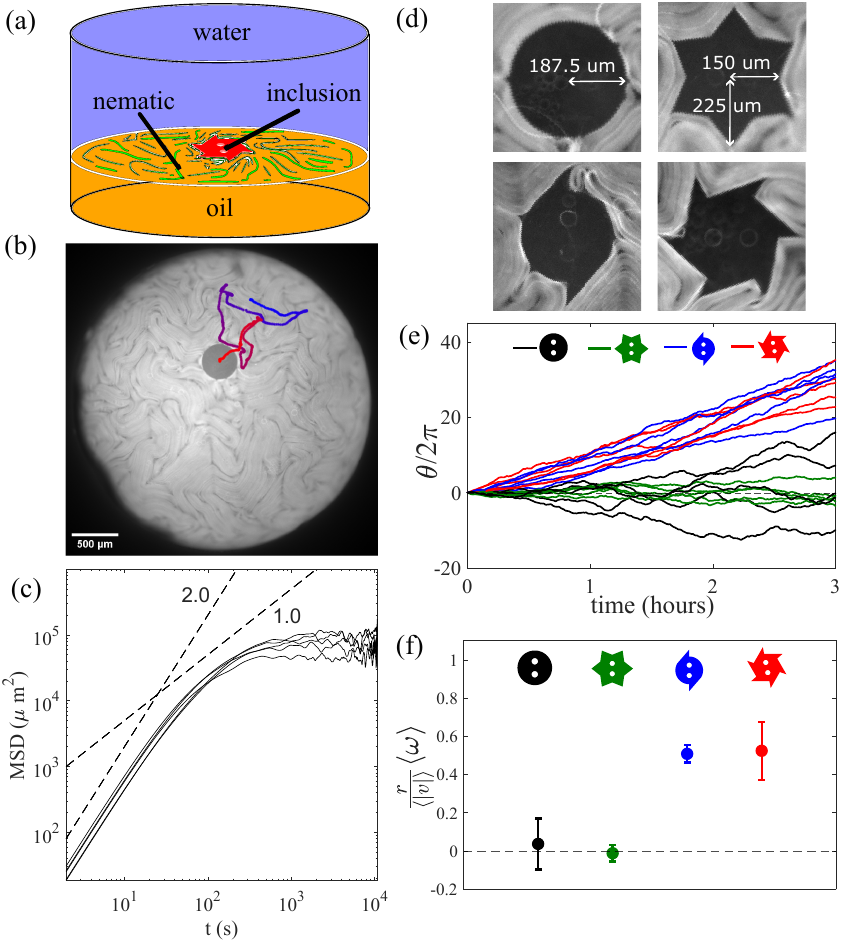}
    \caption{Inclusions in 2D active nematics. (a) An SU-8 inclusion is placed on an oil-water interface, and a MT-kinesin active nematic is formed around it.  (b) An isolated inclusion embedded in the active nematic. The colored line indicates the inclusion trajectory over 17 minutes (t=0 min is blue, and t=17 min is red). (c) MSDs of 5 trajectories of disk-shaped inclusions, each from an independent sample. (d) The four inclusion shapes used in this study. (e) Angular trajectories of different inclusion shapes. Each line corresponds to a single inclusion. For asymmetric gears, $\theta(t)$ trajectories were multiplied by $-1$ for gears with the opposite chirality than the legend cartoon. (f)  Mean rescaled rotation rates of the 4 shapes. The average rotation rate $\langle \omega \rangle$ (rad/s) is rescaled by $\langle \mid v \mid \rangle /r$ ($\mu$m/s) where nematic mean speed is $\langle \abs{v} \rangle$ and inclusion radius is $r$. Error bars indicate standard deviation over 5 independent samples.}
    \label{fig:fig1}
\end{figure}

We developed a method to robustly embed a single particle into a 2D active nematic and observe its dynamics over the sample lifetime [Fig.~\ref{fig:fig1}a]. We first created an oil-water interface inside an open tube of diameter 3.3 mm. The dense oil and lighter aqueous layers were \mytilde 0.4 mm and \mytilde 1.5 mm high, respectively. We fabricated rigid, quasi-2D particles from SU8 photoresist with size of several 100 $\mu$m~\cite{choi2011synthesis}. The open tube setup allowed us to deposit a single particle in each sample [Fig.~\ref{fig:fig1}b], eliminating particle-particle interactions. We then assembled a microtubule (MT)-kinesin active nematic around the particle adsorbed at the oil-water interface~\cite{tayar2022assembling}. The MTs were excluded from the inclusion and aligned parallel to the boundaries. MT bundles sometimes crept under the inclusion, leading to only partial embedding in the nematic layer. To control for variations in the embedding quality, we only quantified  inclusions that were well-embedded in the active nematic layer [Fig. S1]. The imaging setup captured the entire interface, allowing us to track the inclusion as well as the active nematic texture over the sample lifetime. 

We first studied the motion of disk-shaped inclusions of radius $r$=187.5 $\mu$m. Being advected by the chaotic flows the inclusion exhibited a random-walk-like trajectory  [Fig.~\ref{fig:fig1}b]. For the shortest times, the motion was along straight lines and the measured mean-squared displacement (MSD) was ballistic. At longer times ($\approx 45\pm3$) the inclusions changed direction and MSDs transitioned to more diffusive-like behavior. This was followed by a plateau caused by finite-sized confinement [Fig.~\ref{fig:fig1}c]. In comparison, the characteristic timescales of the active nematic Q-tensor and velocity field were $t_Q=13\pm2$ s and $t_v=29\pm4$ s, respectively [Fig. S2]. Furthermore, the mean displacement of the inclusion at the ballistic to diffusive transition time was $86\pm3$ $\mu$m, compared to characteristic length scales of $l_Q=89\pm2$ $\mu$m and $l_v=237\pm4$ $\mu$m for the active nematic.  

Next, we quantified the rotational dynamics of achiral and chiral inclusions including disks, 6-teeth achiral and chiral gears and 2-teeth chiral gears [Fig.~\ref{fig:fig1}d]. All gear-like inclusions had an inner radius $r_i$=150 $\mu$m and outer radius $r_o$=225 $\mu$m (corresponding to the base and tip of the teeth, respectively). Thus, their mean radius matched the disks. The MSDs of these shapes were similar to those of the disk [Fig. S3]. However, their rotational dynamics were distinct. Disk and achiral gears switched regularly between rotations with opposite directions. In comparison, both the chiral 2-teeth and 6-teeth gears rotated persistently in one direction. Intriguingly, the chiral 2-teeth and 6-teeth gears rotated at essentially the same rate. The rotation direction was set by the handedness of the inclusion chirality [Fig.~\ref{fig:fig1}e, Supplementary Videos 1-4]. 

Active nematic samples had significant sample-to-sample variability. To compare data across different samples we rescaled the inclusion's mean rotation rate $\langle \omega \rangle$ by $\langle \abs{v} \rangle/r$, where $\langle \abs{v} \rangle$ is the mean speed of the background nematic averaged over space and time, and $r$ is the inclusion radius (or an average of outer and inner radii, in the case of gears) [Fig.~\ref{fig:fig1}f]. The rescaled rotation rate was \mytilde 0 for disks and achiral gears, and \mytilde 0.5 for both the chiral 2-teeth and 6-teeth gears. Thus, on average, the tangential velocity of the inclusion boundary was about half the speed of the nematic for both chiral 2-teeth and 6-teeth gears. For all inclusions, the instantaneous rotation rate was linearly proportional to the instantaneous mean speed of the active nematic [Fig. S4a, b].

\begin{figure}
    \centering
    \includegraphics[width=\columnwidth]{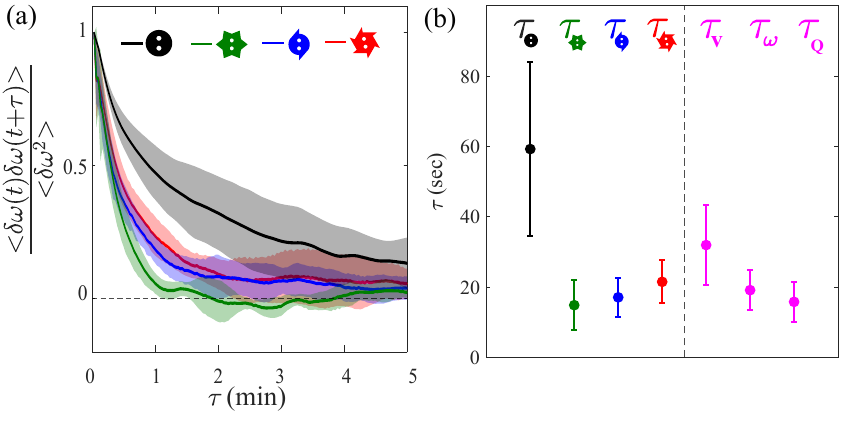}
    \caption{Inclusion shape controls the fluctuations in angular velocity. (a) Auto-correlation of the mean-substracted angular velocity. Shaded regions indicate the standard deviation from different trajectories (n=5). (b) Average fluctuation timescales $\tau_{shape}$ for the different inclusion shapes, and characteristic timescales of bulk active nematic. $\tau_{shape}$ is the lag time at which the autocorrelation reaches 0.5. Similarly, $\tau_v$, $\tau_\omega$, and $\tau_{Q}$ are the lag times at which the velocity, vorticity, and Q-tensor autocorrelation of the background nematic field, respectively, reach 0.5. Error bars indicate standard deviation (n=5).} 
    \label{fig:fig2}
\end{figure}

Both disks and achiral gears failed to show persistent rotation. Nevertheless, they exhibited distinct rotational dynamics. In contrast to gears, disks rotated longer in one direction before switching. This distinction is captured by the temporal autocorrelation of the mean-subtracted angular velocity [Fig.~\ref{fig:fig2}a]. For each shape, we defined a average fluctuation timescale as the lag time at which the autocorrelation dropped to 0.5. The fluctuation timescale for the disks was about three times longer than the gears, which all had timescales close to those of the background active nematic [Fig.~\ref{fig:fig2}b]. Thus, disks stabilized long-lived vortical flows. 

\begin{figure}
    \centering
    \includegraphics[width=\columnwidth]{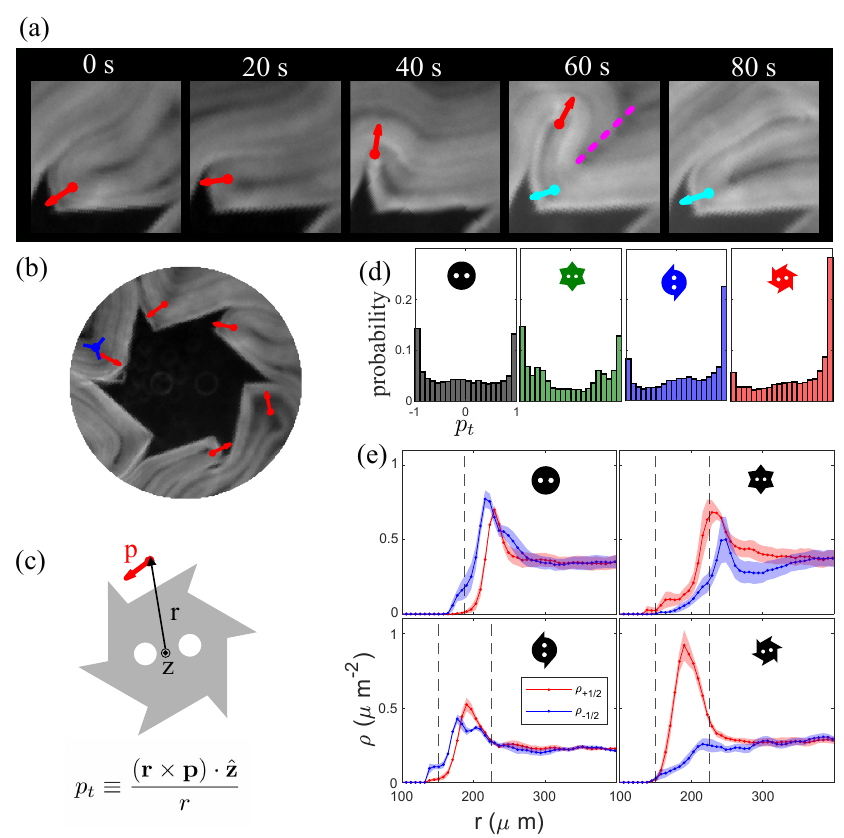}
    \caption{Behavior of topological defects near inclusions. (a) Active nematic defects cycle observed near a corner of the chiral 6-teeth inclusion. The first +1/2 defect is shown in red, the second in cyan. The dashed magenta line indicates a bend deformation causing the nucleation of the second +1/2 defect. (b) Defects detected in the inclusion vicinity. The outer radius of the region is twice the width of the inclusion teeth. +1/2 defects are red, -1/2 defect are blue. (c) Polarity of a +1/2 defect. (d) Distribution of defect polarities near inclusion for the four different shapes. Each histogram corresponds to defects detected during the first 33 minutes of a single sample. (e) Radial distribution of defect densities as a function of distance from the inclusion center. For the disk, the dashed line indicates the disk radius. For the toothed shapes, the first and second dashed lines indicate the inner and outer radii, respectively. The curve thickness indicates the standard error (n=5).}
    \label{fig:fig3}
\end{figure}

To gain insight into the mechanism driving these rotational dynamics, we studied the organization of the active nematic around the inclusions [Supplementary Videos 5-8]. In bulk 2D extensile active nematics, pairs of oppositely charged defects are constantly being created and annihilated ~\cite{Sanchez_Dogic_2012,Giomi2013,Thampi2014}. Due to their asymmetry, the +1/2 defects move in the direction of their polarity. Boundaries influence defect dynamics. In the vicinity of a chiral tooth, +1/2 defects exhibited cyclic dynamics [Fig.~\ref{fig:fig3}a]. At the beginning of each cycle, a +1/2 defect aligned with the long edge of a tooth while pointing against the short edge of the adjacent tooth. Subsequently, as the motile defect approached the corner, it turned away from the inclusion moving along the shorter edge. This motion generated a bend deformation, which nucleated a new defect pair. The +1/2 defect from the newly nucleated pair aligned with the long edge of the tooth, initiating the next cycle. 

To elucidate the relationship between defect organization and inclusion shape, we measured the average polarization of +1/2 defects within a $150$ $\mu$m wide (twice the tooth height) annular region around the inclusion [Fig.~\ref{fig:fig3}b]. We computed the tangential component of the polarity of each +1/2 defect detected in this region, $p_t \equiv \left( \mathbf{r} \times \mathbf{p} \right) \cdot \hat{\mathbf{z}} / r$, with $\mathbf{r}$ the displacement from the inclusion center and $\mathbf{p}$ the defect polarity (unitary length) [Fig.~\ref{fig:fig3}c]. The polarization distributions were strongly skewed towards $p_t=1$ for the chiral gears but were symmetric for the achiral gears and disks [Fig.~\ref{fig:fig3}d]. Since +1/2 defects self-propel in the direction of their polarization, these distributions suggest that defects with $p_t>0$ ($p_t<0$) impart positive (negative) torque to the inclusion, rotating it counterclockwise (clockwise).

In addition to angular polarization, the inclusions also controlled the radial density distribution of topological defects [Fig.~\ref{fig:fig3}e]. The -1/2 defects accumulated at the boundary of disks, with a corresponding buildup of +1/2 defects away from the boundary. In contrast, the chiral 6-teeth gears induced a strong buildup of +1/2 defects, but not -1/2 defects, in the region between the teeth. This is consistent with the previously discussed boundary-induced alignment and nucleation of +1/2 defects. The defect density profile of chiral 2-teeth gears shared characteristics with both the disk and chiral 6-teeth gears. The chiral 2-teeth gears featured both a buildup of -1/2 defects near the boundary away from the teeth and a buildup of +1/2 defects near the teeth. Unlike the chiral 6-teeth gear, the achiral 6-teeth gear induced a buildup of both +1/2 and -1/2 defects just beyond the tips of the teeth. Simply changing the aspect ratio and chirality of the 6 teeth  significantly alters the defect density profile. 

\begin{figure}
    \centering
    \includegraphics[width=\columnwidth]{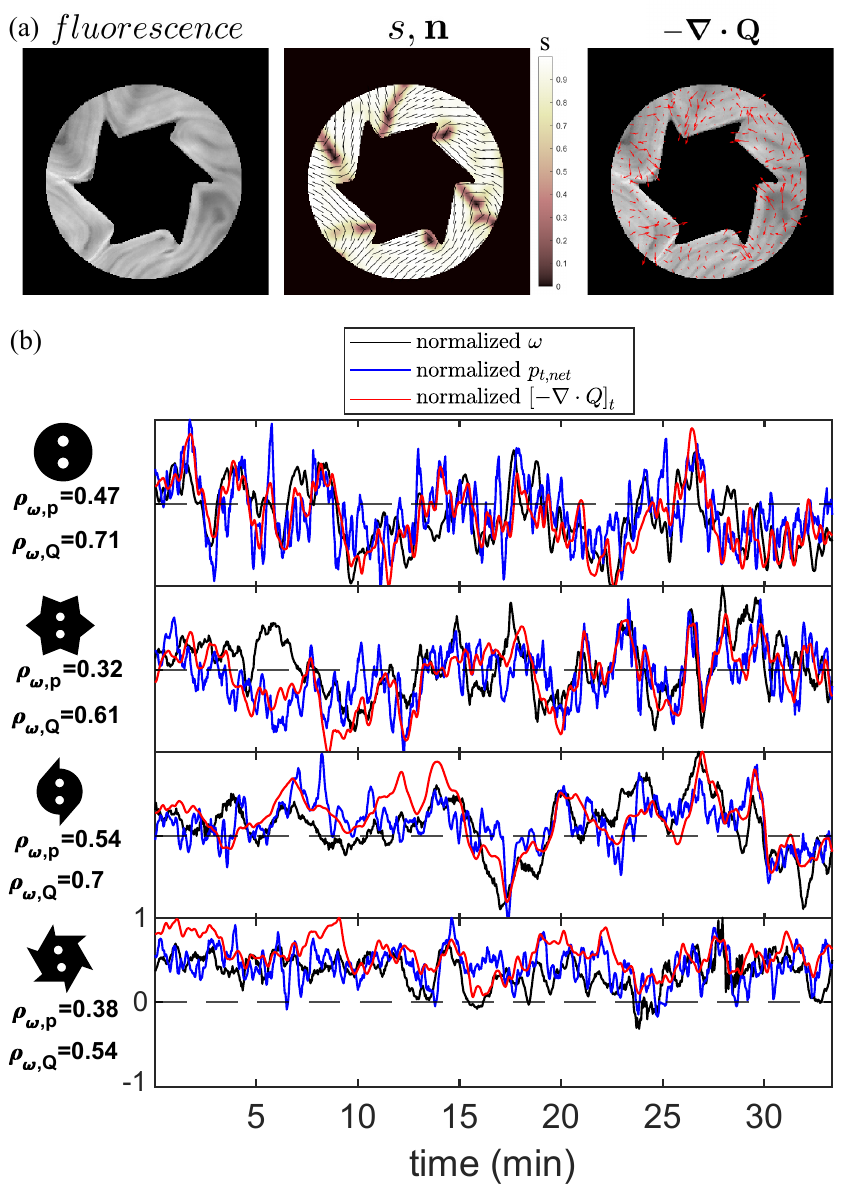}
    \caption{Correlating defect structure to inclusion motion. (a) Computation of $-\mathbf{\nabla \cdot Q}$ from a fluorescence image of MTs. From the image (left), the nematic order parameter $\mathbf{Q}=s(\mathbf{n}\mathbf{n}-\frac{1}{2}\mathbf{1})$ is estimated using the structure tensor. The fields $s$ and $\mathbf{n}$ are displayed (center), as well as $-\mathbf{\nabla \cdot Q}$ (right). (b) Time-series of the normalized angular rotation rate, net +1/2 defect polarization, and the average tangential component of $-\mathbf{\nabla \cdot Q}$ inside an annulus around each of the 4 shapes.} 
    \label{fig:fig4}
\end{figure}

Next, we aimed to correlate inclusion-induced defeat structure to inclusion rotational dynamics. In each frame, we compared the net tangential polarization of all +1/2 defects within the annulus, $p_{t,net}(t)$, to the instantaneous inclusion's rotation rate $\omega(t)$ [Fig.~\ref{fig:fig4}b]. For all shapes,  $\omega(t)$ was weakly correlated with $p_{t,net}(t)$, with a Pearson cross-correlation of $\rho_{\omega,p_t}$ \mytilde 0.2-0.6. Defect polarizations alone may not capture the full coupling of the nematic texture to the inclusion rotation rate, as the deformations of the defect-free regions of the nematic field may impart significant stress to the inclusion. To overcome this issue, we calculated gradients of the nematic order parameter $Q$: 

\begin{equation}
\langle \left[ \div{\mathbf{Q}} \right]_t \rangle_A \equiv -\iint_A \frac{\left( \mathbf{r} \cross \div{\mathbf{Q}} \right) \cdot \mathbf{\hat{z}}}{r} 
\end{equation}

where $A$ denotes an annular region around the inclusion. This quantity is motivated by continuum theories of active nematic hydrodynamics, where active stress is $\mathbf{\sigma}_a = \alpha \mathbf{Q}$. For extensile MT-kinesin active nematics, $\alpha < 0$. The active force field imparted to the fluid is then $\mathbf{f}_a \sim -\div{\mathbf{Q}}$. Thus, $\langle \left[ \div{\mathbf{Q}} \right]_t \rangle_A$ is the average tangential component of the active forces in a region around the inclusion. The nematic order parameter $\mathbf{Q}=s(\mathbf{n}\mathbf{n}-\frac{1}{2}\mathbf{1})$ and its divergence are computed from the fluorescence images [Fig.~\ref{fig:fig4}a]. We found that the $\langle \left[ \div{\mathbf{Q}(t)} \right]_t \rangle_A$ was a more accurate predictor of the rotation rate $\omega(t)$ than $p_{t,net}(t)$, with a cross-correlation of $\rho_{\omega,Q}$ \mytilde 0.4-0.8 for all inclusions [Fig.~\ref{fig:fig4}b]. Active stress $\mathbf{\div{Q}}$ generated away from the inclusion can only influence inclusion motion through the mediating solvent. A complete modeling scheme would require measuring or numerically solving for the fluid flow at the inclusion boundary, and then estimating the active, viscous, and passive elastic stresses. Intriguingly, a quantity involving only the nematic Q-tensor captures the main features of the inclusion's rotational dynamics.

Our experiments demonstrated the rich behavior of inclusions embedded in 2D active nematics. Chiral gear-shaped inclusions rotated persistently in one direction, while achiral disks and gears did not. A previous study observed the persistent unidirectional circulation of an active nematic confined inside a 200 $\mu$m well with a single chiral notch on the boundary \cite{Opathalage2019}. In that case, the notch also served as a site for defect nucleation, although the notch shape and nucleation process were different. Intriguingly, the chiral 2-teeth and 6-teeth gears rotated at the same average rate, with a tangential velocity about half the speed of the bulk active nematic. This contrasts with work where the circulation speed of a 3D isotropic active fluid confined to an annulus decreased with the number of notches on the boundary \cite{Wu_Dogic_2017}. While neither disks nor achiral gears exhibited rectified rotation, they still had distinct dynamics. The angular velocity fluctuations of the disk were correlated over longer timescales than those of any of the gears. These results suggest a nontrivial relationship between inclusion shape and rotational dynamics which has yet to be systematically explored. From an entirely different perspective, these experiments extend previous work on mixtures of equilibrium liquid crystals and conventional colloids, which demonstrated control of the spatial structure and the topology of defects \cite{Poulin_Lubensky_Weitz_1997, Gu_Abbott_2000, Musevic_Zumer_2006, Martinez_Smalyukh_2012}. Creating active liquid crystal/colloid composites offers a tantalizing possibility of using inclusions in \textit{active} nematics to control both the spatial and temporal dynamics of motile topological defects which in turn affect the colloidal motion\cite{Rajabi_Lavrentovich_2021}. 

Our system has both similarities and differences with studies of gears powered by motile bacteria \cite{Sokolov2010, DiLeonardo2010}. In both cases, chiral gears rotated unidirectionally, while achiral gears did not. As polar swimmers, individual bacteria were more likely to get trapped along the long edge while pointing towards the short edge. The bacteria exerted force by direct contact on the short edges, leading to a net torque. While built from apolar constituents, active nematics exhibit +1/2 topological defects which are polar quasi-particles. Moreover, the dynamics of +1/2 defects near the chiral teeth [Fig.~\ref{fig:fig3}a] suggest that they play an analogous role to the swimming bacteria, pushing more against the short edge than the long edge and therefore transferring net torque to the gear. A key difference, however, is that defects are continuously created and annihilated within an active nematic, and this regeneration is boundary controlled. The presence of the chiral teeth induces the local cyclic nucleation of +1/2 defects [Fig.~\ref{fig:fig3}a], rather than capturing defects from the bulk. Unlike bacteria, forces exerted by the defects on the inclusion cannot come from direct contact. The active, elastic, and viscous stresses felt at the inclusion boundary are caused by the hydrodynamic flows and associated nematic texture. Our experiments demonstrate a need to understand the effective interaction between defects and inclusion boundaries in active nematics. 

In conclusion, we studied the dynamics of inclusions embedded in 2D active nematics with parallel anchoring. Chiral gear-shaped inclusions rotated persistently in one direction, while disk and achiral gear had unbiased rotation. Inclusion shape influenced the polarization and distribution of nearby topological defects. These results show that inclusion motion in active nematics arises from a delicate interplay between the inclusion shape, organization of the surrounding nematic, and the boundary-induced self-organized active stresses. 

We acknowledge useful discussions with Cristina Marchetti and Zhihong You. Work was supported by the US Department of Energy, Office of Basic Energy Sciences, through award DE-SC0019733.

\bibliographystyle{apsrev4-2}
\bibliography{main}

\end{document}